\documentclass[prl,twocolumn,superscriptaddress]{revtex4-1}

\usepackage{graphicx}  
\usepackage{dcolumn}   
\usepackage{bm}        
\usepackage{amssymb}   
\usepackage{verbatim}
\usepackage{amsmath}
\usepackage{setspace}

\graphicspath{{figures/}}

\begin{document}

\widetext

\title{Optimizing a Dynamical Decoupling Protocol for Solid-State Electronic Spin Ensembles in Diamond}

\author{D. Farfurnik}
\affiliation{Racah Institute of Physics, Hebrew University, Jerusalem 9190401, Israel}
\affiliation{The Center for Nanoscience and Nanotechnology, Hebrew University, Jerusalem 9190401, Israel}

\author{A. Jarmola}
\affiliation{Department of Physics, University of California, Berkeley, California 94720-7300, USA}

\author{L. M. Pham}
\affiliation{Harvard-Smithsonian Center for Astrophysics, Cambridge, MA 02138, USA}

\author{Z. H. Wang}
\affiliation{Department of Chemistry, University of Southern California, Los Angeles, California 90089, USA}

\author{V. V. Dobrovitski}
\affiliation{Ames Laboratory, Iowa State University, Ames, Iowa 50011, USA}

\author{R. L. Walsworth}
\affiliation{Harvard-Smithsonian Center for Astrophysics, Cambridge, MA 02138, USA}
\affiliation{Department of Physics, Harvard University, Cambridge, Massachusetts 02138, USA}

\author{D. Budker}
\affiliation{Helmholtz Institute, JGU, Mainz, Germany}
\affiliation{Department of Physics, University of California, Berkeley, California 94720-7300, USA}

\author{N. Bar-Gill}
\affiliation{Racah Institute of Physics, Hebrew University, Jerusalem 9190401, Israel}
\affiliation{The Center for Nanoscience and Nanotechnology, Hebrew University, Jerusalem 9190401, Israel}
\affiliation{Department of Applied Physics, Rachel and Selim School of Engineering, Hebrew University, Jerusalem 9190401, Israel}

\date{\today}

\begin{abstract}

We demonstrate significant improvements of the spin coherence time of a dense ensemble of nitrogen-vacancy (NV) centers in diamond through optimized dynamical decoupling (DD). Cooling the sample down to $77$ K suppresses longitudinal spin relaxation $T_1$ effects and DD microwave pulses are used to increase the transverse coherence time $T_2$ from $\sim 0.7$ ms up to $\sim 30$ ms. We extend previous work of single-axis (CPMG) DD towards the preservation of arbitrary spin states. Following a theoretical and experimental characterization of pulse and detuning errors, we compare the performance of various DD protocols. We identify that the optimal control scheme for preserving an arbitrary spin state is a recursive protocol, the concatenated version of the XY8 pulse sequence. The improved spin coherence might have an immediate impact on improvements of the sensitivities of AC magnetometry. Moreover, the protocol can be used on denser diamond samples to increase coherence times up to NV-NV interaction time scales, a major step towards the creation of quantum collective NV spin states.

\end{abstract}

\maketitle


In recent years, atomic defects in diamond have been the subject of a rapidly growing area of research. The most well-studied of these diamond defects is the nitrogen-vacancy (NV) color center, whose unique spin and optical properties make it a leading candidate platform for implementing magnetic sensors \cite{Taylor2008, Maze2008, Balasubramanian2008, Grinolds2011, Pham2011, Pham2012, Acosta2009, Acosta2010,DeLange2011,Mamin2014} as well as qubits, the building blocks for applications in quantum information.
In particular, NV spin coherence times longer than a millisecond have been achieved in single NV centers at room temperature, either through careful engineering of a low spin impurity environment during diamond synthesis \cite{Balasubramanian2009} or through application of pulsed \cite{Ryan2010,Naydenov2011,Shim2012,Toyli2013} and continuous \cite{Hirose2012,Cai2012} dynamical decoupling (DD) protocols. These long single NV spin coherence times have been instrumental in demonstrating very sensitive magnetic \cite{Taylor2008, Maze2008, Balasubramanian2008, Grinolds2011, Pham2011, Pham2012, Acosta2009, Acosta2010,DeLange2011,Mamin2014}, electric \cite{Dolde2011}, and thermal \cite{Toyli2013} measurements as well as high-fidelity quantum operations \cite{Bernien2013,Tsukanov2013}.

Achieving similarly long spin coherence times in ensembles of NV centers can further improve magnetic sensitivity \cite{Pham2011,Pham2012} and, moreover, may open up new avenues for studying many-body quantum entanglement. For example, achieving NV ensemble spin coherence times longer than the NV-NV interaction timescales within the ensemble could allow for the creation of non-classical spin states \cite{Cappellaro2009,Bennett2013,Weimer2013}. Recently, NV ensemble spin coherence times up to $\sim 600$ ms have been demonstrated by performing Carr-Purcell-Meiboom-Gill (CPMG) DD sequences at lower temperatures to reduce phonon-induced decoherence \cite{BarGill2013}. The CPMG sequence preserves only a single spin component efficiently, however; experimentally, in the presence of pulse imperfections, the CPMG DD protocol cannot protect a general quantum state \cite{DeLange2010,Wang2012a,Wang2012}, as is necessary for applications in quantum information and sensing. To date, the preservation of arbitrary NV spin states has been considered only
in a limited fashion, mostly at room temperatures and for single NV centers \cite{Ryan2010,Naydenov2011,Shim2012}. However, no fundamental study yet considered the robustness of various DD protocols on NV ensembles. \emph{In this work, we perform a theoretical and experimental analysis of the performance of several DD protocols, including standard CPMG and XY-based pulse sequences as well as modifications thereon,
and extract an optimized protocol for preserving a general NV ensemble state at $77$ K}. We observe an extension of the arbitrary NV ensemble state from a coherence time $\sim 0.7$ ms of an Hahn-Echo measurement up to a coherence time $\sim 30$ ms, which is more than an order of magnitude improvement. Although higher coherence times were demostrated for preserving a specific spin state \cite{BarGill2013}, in this work we fundamentally study and optimize a DD protocol for preserving an arbitrary state.


\begin{figure}[t]
		\includegraphics[width=0.85\columnwidth]{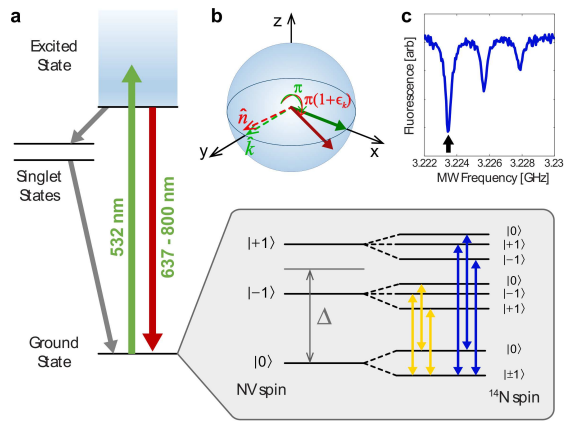} 
        \caption{(a) Energy levels of the negatively charged NV center, including the $^{14}$N hyperfine splitting, $\Delta$ is the zero-field splitting. (b) Bloch sphere diagram illustrating the two main types of pulse imperfection: $\epsilon_{\hat{k}}$ represents the deviation from an ideal rotation angle $\pi$, and $\hat{n} = (n_x, n_y, n_z)$ is the actual rotation axis, which can deviate from $\hat{k} = (k_x, k_y, 0)$. (c) Optically detected magnetic resonance measurement of $|0\rangle \leftrightarrow |+1\rangle$ transition in an NV ensemble. Hyperfine interactions between the NV electronic and the $^{14}$N nuclear spins form three NV resonances, and a strong static field $\sim 300$ G polarizes the $^{14}$N nuclear spins into the $|-1\rangle$ spin state.}
        \label{fig:nvstructure}
\end{figure}

The NV center is composed of a substitutional nitrogen atom (N) and a vacancy (V) on adjacent lattice sites in the diamond crystal.
The electronic structure of the negatively charged NV center has a spin-triplet ground state, where the $m_s=\pm 1$ sublevels experience a zero-field splitting ($\sim 2.87$ GHz) from the $m_s = 0$ sublevel due to spin-spin interactions [Fig. \ref{fig:nvstructure}(a)]. Application of an external static magnetic field along the NV symmetry axis Zeeman shifts the $m_s=\pm 1$ levels and allows one to treat the $m_s= 0,+1$ spin manifold (for example) as an effective two-level system. The NV spin state can be initialized in the $m_s= 0$ state with off-resonant laser excitation, coherently manipulated with resonant microwave (MW) pulses, and read out optically via spin-state-dependent fluorescence intensity of the phonon sideband \cite{Taylor2008}.

The NV spin bath environment is typically dominated by $^{13}$C nuclear and N paramagnetic spin impurities, randomly distributed in the diamond crystal. These spin impurities create different time-varying local magnetic fields at each NV spin, which can be approximated as a random local magnetic field that fluctuates on a timescale set by the mean interaction between spins in the bath. This random field induces dephasing of freely precessing NV spins on a timescale $T_2^*$ \cite{Pham2012,Acosta2009,deSousa2009,BarGill2012}.
Dynamical decoupling pulse sequences can suppress the effect of the spin bath noise and thus preserve the NV spin coherence up to a characteristic time $T_2$ \cite{BarGill2013,BarGill2012}. In the ideal case of perfect pulses, various DD protocols (e.g., CPMG, XY, etc.) are equally effective at preserving an arbitrary NV ensemble spin state. Experimentally, however,  off-resonant driving due to the NV hyperfine structure \cite{Suppl} and other pulse imperfections significantly affect the performance of individual DD protocols. In order to overcome these pulse imperfections, we optimize a DD protocol for an ensemble of NV spins. 

\begin{figure}[!t]
  \includegraphics[width=1\columnwidth]{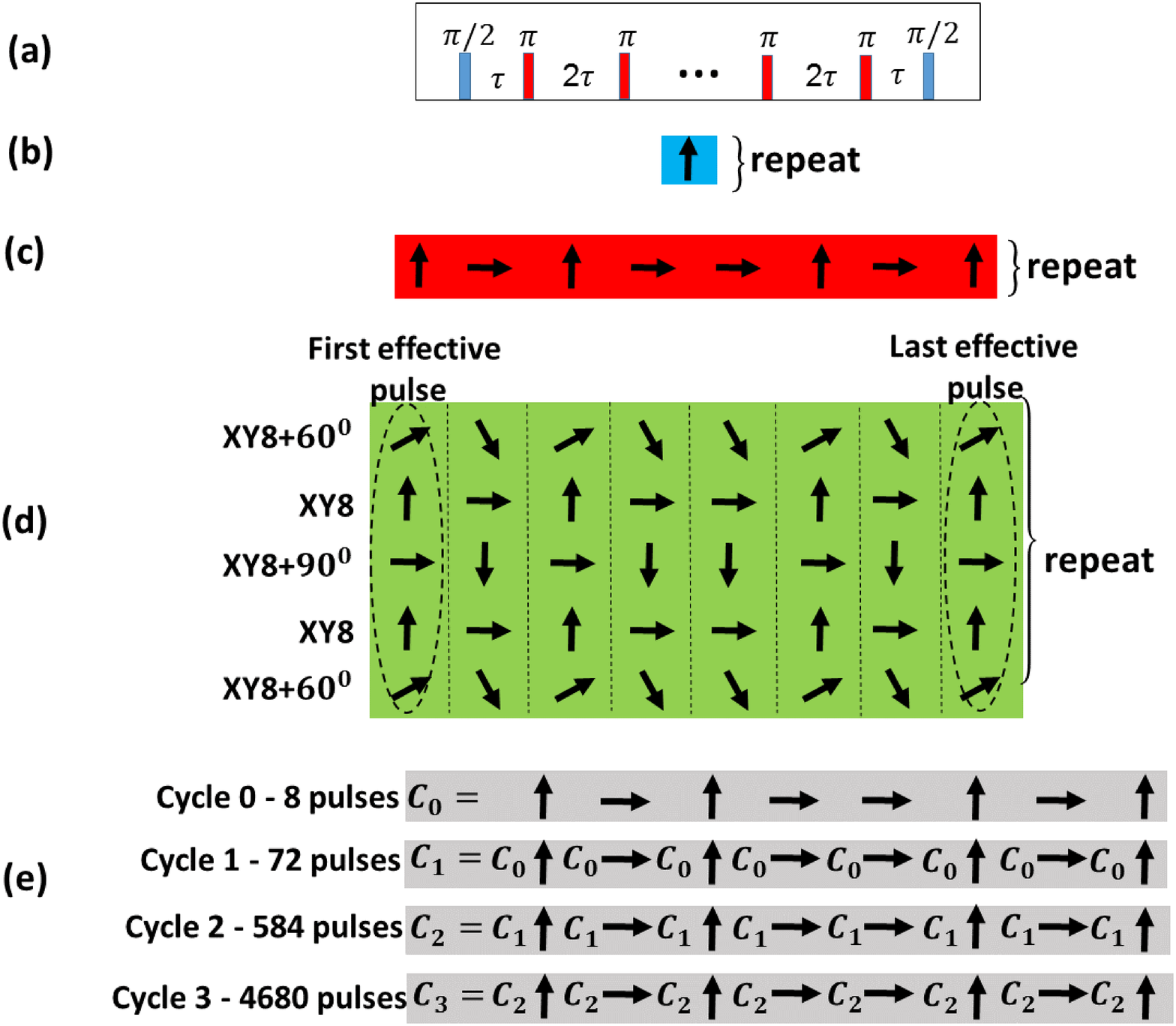} 
  \caption{Dynamical decoupling protocols.  The directions of the arrows in the scheme represent the phases of the pulses. For each sequence, the free evolution time between pulses $2\tau$ was swept to obtain a full coherence curve. (a) General DD scheme. (b) CPMG. (c) XY8. (d) KDD version of XY8: each ($\pi$)-pulse from an XY8 sequence is replaced by five adjacent ($\pi$)-pulses, with additional phases of $(\pi)_{60^{\circ}} - (\pi)_{0^{\circ}} - (\pi)_{90^{\circ}}  - (\pi)_{0^{\circ}} - (\pi)_{60^{\circ}}$, keeping a free evolution time of $2\tau$ between them. (e) Concatenated version of XY8: the first applied cycle (cycle 0) is a single conventional XY8. Each of the following cycles is constructed recursively from the previous ones: eight pulses of conventional XY8 are always applied, but between every two of them, the whole cycle from the previous iteration is applied.}
  \label{fig:ddsequences}
\end{figure}

Figure \ref{fig:ddsequences}(a) illustrates the general structure of the DD protocols explored in this work.
In each protocol, $(\pi)$-pulses about a rotation axis determined by the specific DD protocol are applied, with a free evolution interval of time $2 \tau$ between them.
In the regime where the pulse durations are short compared to the free evolution interval between adjacent pulses,
each pulse can be expressed in terms of a spin rotation operator \cite{Wang2012a,Wang2012}
\begin{equation}
U_{\hat{k}}=\exp{\{-i\pi(1+\epsilon_{\hat{k}})[\vec{S}\cdot \hat{n} ]\}}. \label{eq:rot1}
\end{equation}

Equation \eqref{eq:rot1} incorporates the two main types of pulse imperfection: $\epsilon_{\hat{k}}$ represents the deviation from an ideal rotation angle $\pi$, and $\hat{n} = (n_x, n_y, n_z)$ is the actual rotation axis, which can deviate from $\hat{k} = (k_x, k_y, 0)$ [Fig. \ref{fig:nvstructure}(b)]. Generally, imperfections in the rotation angle ($\epsilon_{\hat{k}}$) may be caused by limitations in pulse timing resolution and amplitude stability of the MW field source, as well as static and MW field inhomogeneity over the measurement volume; and imperfections in the rotation axis may be caused by phase instability in the MW field source. In addition to general experimental pulse errors, the specific physical system of the NV spin ensemble introduces additional pulse imperfections. Most notably, hyperfine interactions between the $^{14}$N nuclear spin ($I = 1$) of the NV center and the NV electronic spin result in three transitions each separated by $\sim 2.2$ MHz in the, e.g., NV $m_s = 0 \leftrightarrow +1$ resonance \cite{Jelezko2006} [Fig. \ref{fig:nvstructure}(c)].

The total evolution operator of a general DD sequence containing $n$ $(\pi)$-pulses can then be expressed as
\begin{equation}
U_{\rm{DD}}=U_d(\tau) \cdot U_{\hat{k}_n} \cdot U_d(2\tau) \cdot U_{\hat{k}_{n-1}} \cdot U_d(2\tau) \cdot... \cdot U_d(2\tau) \cdot U_{\hat{k}_1} \cdot U_d(\tau), \label{eq:dd}
\end{equation}
where $U_d$ is the free evolution operator. It is clear that without compensation for pulse imperfections in the spin rotation operators, accumulating errors will result in a severe loss of coherence even in the limit of free evolution time $\tau \rightarrow 0$.
First, we study the robustness of conventional CPMG and XY-based DD protocols, summarized in Figure \ref{fig:ddsequences} (b) (c), in order to determine which protocol is the most robust against pulse imperfections caused by general experimental limitations as well as those specific to NV ensembles. Realizing that enhanced robustness is necessary, we reduce the effects of the imperfections by optimizing experimental parameters (see detailed experimental setup description below) and modify the basic XY sequences by introducing pulses with additional phases [Fig. \ref{fig:ddsequences}(d)] and concatenated cycles [Fig. \ref{fig:ddsequences}(e)].
Similar DD protocol optimization has been performed in the past for phosphorus donors in silicon \cite{Wang2012a} and single NV centers \cite{Ryan2010,DeLange2010,Wang2012,Souza2011}.

In the conventional CPMG DD protocol \cite{Meiboom1958}, all ($\pi$)-pulses are applied along the same axis ($x$) [Fig. \ref{fig:ddsequences}(b)]; consequently, only coherence along one spin component is well-preserved. The XY family of DD protocols \cite{Gullion1990} applies pulses along two perpendicular axes ($x,y$) in order to better preserve spin components along both axes equally [Fig. \ref{fig:ddsequences}(c)]. We also explored two DD protocols which introduce additional modifications on the basic XY pulse sequences in order to improve its robustness against pulse errors. The first modification, the Knill Dynamical Decoupling (KDD) pulse sequence \cite{Ryan2010, Souza2011}, introduces additional phases, thereby symmetrizing the XY-plane further and reducing the effects of pulse errors due to off-resonant driving and imperfect $\pi$-flips. In the KDD protocol, each $(\pi)$-pulse in a basic XY sequence is replaced by five pulses with additional phases given by $(\pi)_{60^{\circ}} - (\pi)_{0^{\circ}} - (\pi)_{90^{\circ}}  - (\pi)_{0^{\circ}} - (\pi)_{60^{\circ}}$, where the $2\tau$ free evolution interval between adjacent $(\pi)$-pulses timing is preserved [Fig. \ref{fig:ddsequences}(d)].
The second modification employs concatenation, a recursive process in which every cycle is constructed from the previous cycles [Fig. \ref{fig:ddsequences}(e)], and each level of concatenation corrects higher orders of pulse errors \cite{Khodjasteh2005,Witzel2007}.


We performed measurements on an isotopically pure ($99.99\%$ $^{12}$C) diamond sample with N concentration $\sim 2 \times 10^{17}$ cm$^{-3}$ and NV concentration $\sim 4 \times 10^{14}$ cm$^{-3}$ (Element Six), grown via chemical vapor deposition.
The sample was placed in a continuous flow cryostat (Janis ST-500) and cooled with liquid nitrogen to $77$ K, significantly reducing phonon-related decoherence to allow for NV spin coherence times $\gg 1$ ms \cite{BarGill2013,Jarmola2012}. A 532-nm laser optically excited an ensemble of $\sim 10^4$ NV centers within a $\sim 25\ \mu$m$^3$ measurement volume, and the resulting fluorescence was measured with a single photon counting module. A permanent magnet produced a static magnetic field $B_0 \sim 300$ G along the NV symmetry axis, Zeeman splitting the $m_s = \pm1$ spin sublevels. To coherently manipulate the NV ensemble spin state, we used a 70-$\mu$m diameter wire to apply a MW field resonant with the $m_s = 0 \leftrightarrow +1$ transition. The spin rotation axes of the individual DD pulses were set through IQ modulation of the MW carrier signal from the signal generator (SRS SG384).

As discussed previously, one of the sources of pulse imperfections for NV centers is the hyperfine structure in the NV resonance spectrum; specifically, resonant driving of one of the hyperfine transitions results in detuned driving of the other two, introducing both spin rotation angle and spin rotation axis errors.
We mitigate these effects by: (i) applying a strong static magnetic field ($\sim300$ G) to polarize the $^{14}$N nuclear spins \cite{Fischer2013} into one hyperfine state which we drive [Fig. \ref{fig:nvstructure}(c)] and (ii) applying a strong MW field to drive the NV transition with Rabi frequency ($\sim15$ MHz) much greater than the detuning due to NV hyperfine splitting ($\sim2.2$ MHz). Furthermore, we minimize general experimental pulse errors due to pulse timing and amplitude imperfections, MW carrier signal phase imperfections, and static and MW field inhomogeneities over the measurement volume \cite{Suppl}. We estimate that the pulse imperfections remaining after this optimization are characterized by $\epsilon_{\hat{k}} \approx 0.15$ and $n_z \approx 0.25$.

\begin{figure}[!t]	
  \includegraphics[width=1.05\columnwidth]{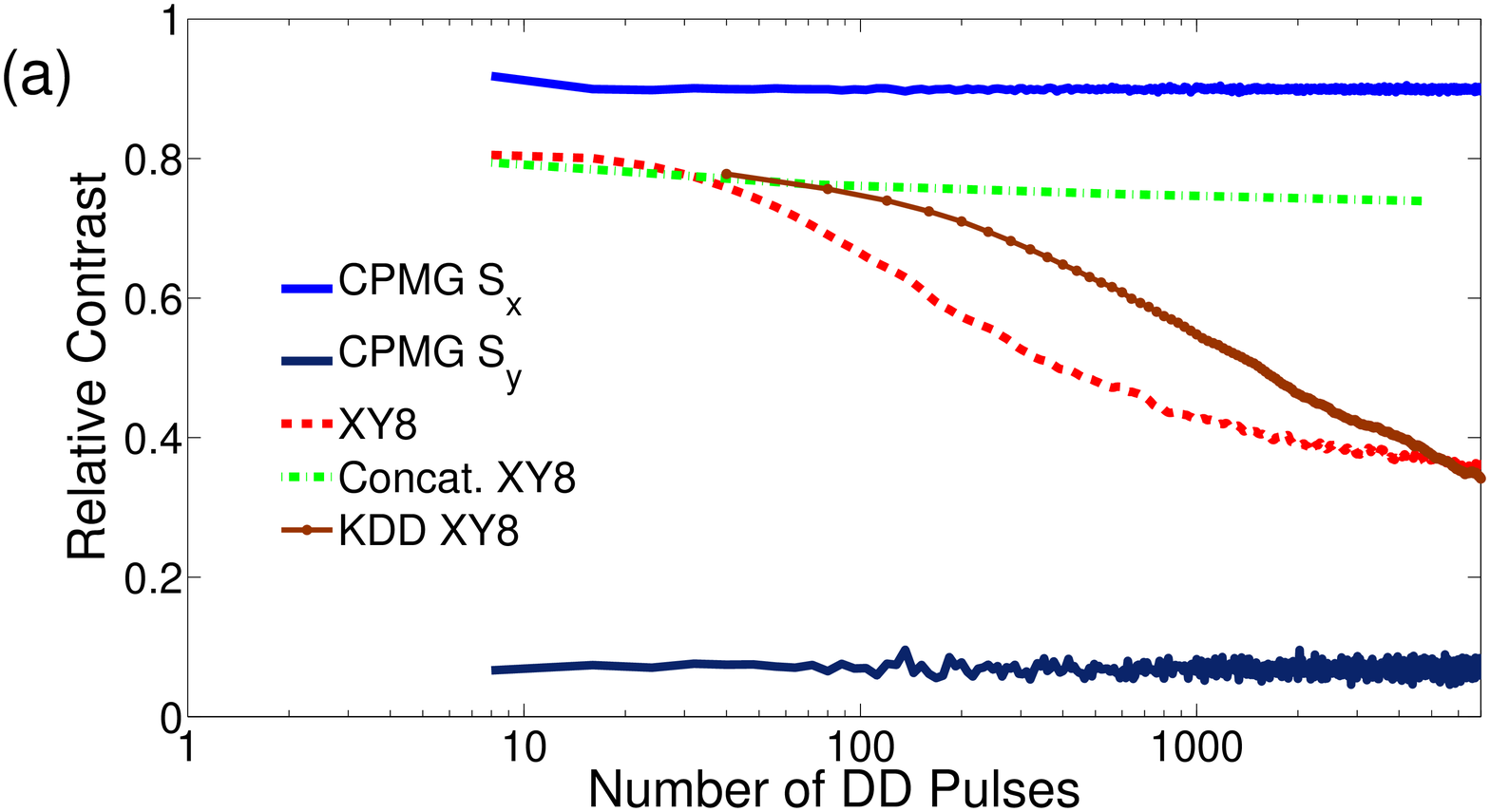} 
  \includegraphics[width=1.05\columnwidth]{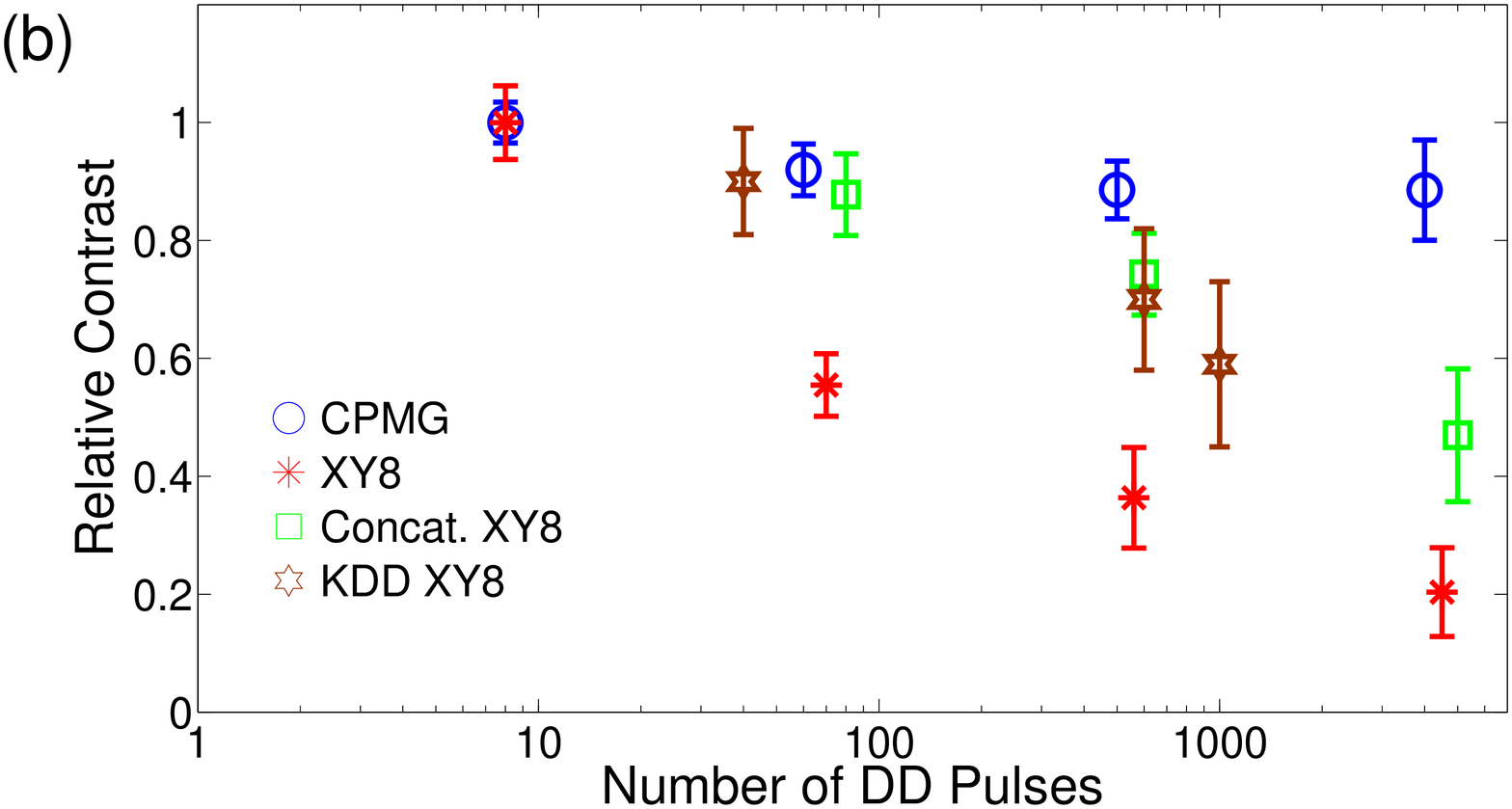} 
  \caption{Relative contrast in the decoherence-free limit ($\tau \ll \frac{T_2}{n}$) of DD protocols as a function of number of pulses. For clarity purposes, the simulation is separated from the experimental results. (a) Simulation of the effect of non-ideal $(\pi)$-pulses according to Equation \eqref{eq:dd}. All XY8-based sequences performed similarly for initialization at $S_x$ and $S_y$. (b) Experimental results. The relative contrast is determined via normalizing with a Hahn-Echo measurement in the decoherence-free limit. At the perpendicular axis, the contrast of XY-based sequences is similar, but the CPMG contrast vanishes completely, as demonstrated in the supplemental material \cite{Suppl}}
  \label{fig:contrastvn}
\end{figure}

In order to determine how well each of the four DD protocols preserves a general NV ensemble spin state, we measure the NV spin coherence of two orthogonal initial spin components $S_x$ and $S_y$. The $S_x$ spin component is prepared and measured by applying the initial and final $(\pi/2)$-pulses about the $y$ axis; likewise, the $S_y$ spin component is prepared and measured by applying the initial and final $(\pi/2)$-pulses about the $x$ axis. We first characterize the robustness of each DD protocol against pulse imperfections by measuring NV ensemble spin coherence in the short free evolution (i.e., decoherence-free) limit $2n\tau \ll T_2$ (while remaining in the regime of infinitely narrow MW pulses) and normalizing against the NV ensemble spin coherence of a 1-pulse Hahn-Echo measurement in the same limit. We plot the experimental results in Figure \ref{fig:contrastvn}(b) for each of the DD protocols as a function of number of pulses $n$, where a relative contrast of 1 corresponds to perfect preservation of NV ensemble spin coherence and relative contrast of 0 corresponds to a mixed state. Incorporating estimated pulse imperfection values into Equations \eqref{eq:rot1} and \eqref{eq:dd}, we also plot simulated relative contrast of each DD protocol as a function of number of pulses [Fig. \ref{fig:contrastvn}(a)].

The CPMG protocol maintains the highest relative contrast for the spin component along the spin rotation axis of the DD pulses ($S_x$) but the lowest relative contrast for the spin component along the perpendicular axis ($S_y$) \cite{Suppl} , as expected.
The relative contrast of XY-based sequences is comparable for both spin components \cite{Suppl} but drops as the number of pulses increases, indicating that while the XY-based protocol is able to symmetrically compensate for pulse errors and thus preserve an arbitrary NV ensemble spin state, accumulating pulse errors due to imperfect compensation eventually limit the sequence to $\sim500$ pulses. Within the XY family, we compared XY4, XY8, and XY16 pulse sequences \cite{Gullion1990} and found XY8 to show the best performance \cite{Suppl}.
The KDD protocol, which introduces more spin rotation axes to further symmetrize pulse error compensation, and the concatenated protocol, which constructs the pulse sequences recursively in order to correct for higher orders of pulse errors both improve upon the conventional XY8 sequence, maintaining higher relative contrast for both spin components to $>500$ pulses. Note that the measurements are in qualitative agreement with the simulations. Quantitavely, however, there is a disagreement, and the experimental results for the relative contrast are slightly lower than the simulation suggests. In particular, the contrst of the concatenated XY8 protocol does not change with the number of pulses according to the simulation, which disagrees with the experimental data. This disagreement is likely caused by the interplay between pulse errors and decoherence effects, which was not taken into account in the simulation and will be the subject of a future research.

\begin{figure}[!ht]
        \centering
		\includegraphics[width=0.95\columnwidth]{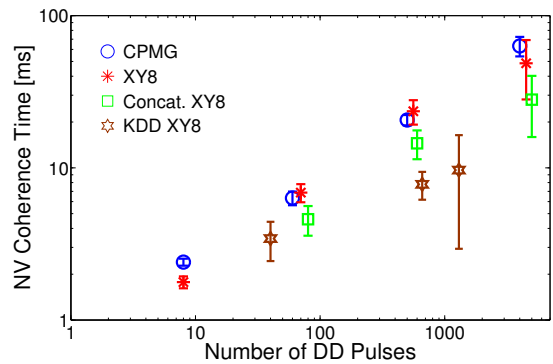}
        \caption{Experimental results of the coherence time of DD sequences as a function of the number of pulses, after initialization at $S_x$. The results after initialization at $S_y$ are shown in the supplemental material \cite{Suppl}}
\label{fig:T2vn}
\end{figure}

The measured NV ensemble spin coherence time is plotted as a function the number of pulses for each DD protocol in Figure \ref{fig:T2vn}. The CPMG, XY8, and concatenated XY8 protocols all extend the NV spin coherence time as expected, given the nitrogen-impurity-dominated spin bath environment \cite{BarGill2012}. However, the KDD protocol is less effective at extending the NV spin coherence time; this underperformance is probably due to the fact that the phase difference between adjacent pulses in KDD (sometimes $60^{\circ}$) is smaller than in other sequences ($90^{\circ}$), making phase errors more significant \cite{Suppl}. 

In conclusion, after optimizing experimental parameters to minimize pulse imperfections,
we found the most robust DD protocol for preserving an arbitrary spin state in an NV ensemble system to be the concatenated XY8 pulse sequence. By compensating for higher order pulse errors, the concatenated XY8 sequence maintains higher relative contrast than the conventional XY8 sequence and is expected to ultimately outperform the KDD sequence for larger numbers of pulses. Furthermore, the concatenated XY8 sequence achieves longer NV ensemble spin coherence times than the KDD sequence. At $77$ K, we measured an extension of the arbitrary spin state of an ensemble of $\sim10^4$ NV centers by a factor of $\sim 40$ and up to $\sim30$ ms.

The optimized DD protocol determined in this work may already have an immediate impact in improving the sensitivity of NV magnetometry \cite{Pham2012} and, moreover, may be useful for quantum information applications. The sample in this work has nitrogen density $\sim 2 \times 10^{17}$ cm$^{-3}$ and NV density $\sim 4 \times 10^{14}$ cm$^{-3}$, corresponding to N-to-NV conversion efficiency $\sim0.2\%$ and typical NV-NV interaction time $\sim 150$ ms. Using standard sample processing techniques, such as electron irradiation \cite{Acosta2009}, to modestly improve the N-to-NV conversion efficiency to $\sim1\%$, the concatenated XY8 pulse sequence can increase the NV ensemble spin coherence time to the NV-NV interaction time. In such a case, MREV-based techniques \cite{Mansfield1973} can be applied to average out the NV-NV interactions and introduce effective Hamiltonians \cite{Cappellaro2009,Bennett2013,Weimer2013}, thereby creating self engineered quantum states (e.g. squeezed states) in NV ensemble systems.

\begin{acknowledgments}

We thank Gonzalo A. \'{A}lvarez for fruitful discussions.
This work has been supported in part by the EU CIG, the Minerva ARCHES award, the Israel Science Foundation (grant No. 750/14), the Ministry of Science and Technology, Israel, the CAMBR fellowship for Nanoscience and Nanotechnology, the Binational Science Foundation Rahamimoff travel grant, the German-Israeli Project Cooperation (DIP) program, the NSF through (grant No. ECCS-1202258), and the AFOSR/DARPA QuASAR program. Work at the Ames Laboratory was supported by the Department of Energy - Basic Energy Sciences under Contract No. DE-AC02-07CH11358.
\end{acknowledgments}

\bibliography{nvbibliography}

\end{document}